\documentclass{aa}

\def\mkm{{\mu}\rm{m}}
\def\bet{{\beta}~\rm Pic}
\def\be{\begin{equation}}
\def\ee{\end{equation}}
\def\bea{\begin{eqnarray}}
\def\eea{\end{eqnarray}}
\def\bc{\begin{center}}
\def\ec{\end{center}}
\newcommand{\bi}{\begin{itemize}}
\newcommand{\ei}{\end{itemize}}

\usepackage{graphics}
\begin{document}
\thesaurus{06     
           (08.09.2  $\beta$ Pictoris; 
            08.03.4; 
            09.04.1; 
            02.16.2) 
           }

\title{Circumstellar disc of $\beta$ Pictoris:
constraints on grain properties from polarization}

\author{N.V.~Voshchinnikov\inst{1}
\and    E.~Kr\"ugel\inst{2}}

\institute{ Sobolev Astronomical Institute, St.~Petersburg University,
            Bibliotechnaya pl.~2,
            St.~Petersburg--Peterhof, 198904 Russia
\and        Max--Planck--Institut f\"ur Radioastronomie,
            Auf dem H\"ugel 69, D--53121 Bonn, Germany}

\offprints{Nikolai V. Voshchinnikov, e--mail: nvv@aispbu.spb.su}
\date{Received $<$date$>$; accepted $<$date$>$}
\authorrunning{N.V.~Voshchinnikov \& E.~Kr\"ugel}
\titlerunning{Modelling of $\bet$ polarization}
\maketitle

\begin{abstract}
We model the positional dependence of the optical polarization (BVRI--bands) in
the circumstellar disc of $\beta$ Pictoris as observed by
Gledhill et al.~(\cite{gledh}) and Wolstencroft et al.~(\cite{wo95}).
The particles are spherical, have a size
distribution $n(a)\sim a^{-q}$ and their number density decreases
with distance from the star as $r^{-s}$.
We consider both compact and porous grains.
Varying the grain size and the exponent, $s$, of the density distribution as
well as the refractive index of the grain material $m$, we find that the
measured polarization, colours and brightness distribution in the disc are best
reproduced by a model in which the grains are larger than interstellar grains
(the minimum grain size $a_{\rm min}=0.15\,\mu$m).  The value of the maximum size
is ill determined because it has little influence and was taken to be
$a_{\rm max}=100\,\mu$m.  The best--fit exponents of the power
laws are $q=3.2$ and $s=3$.
The grains have in the R--band a refractive index $m_R=1.152-0.005i$.  Such a
value is roughly appropriate for porous grains where half of the volume is ice
and the other half vacuum, or where 24\% of the volume consists of silicate and
the remaining 76\% of vacuum.

\end{abstract}

\section{Introduction}
In the search for extrasolar planetary systems, $\beta$ Pictoris and other
Vega--type stars have become very popular and our present knowledge
is summarized in several reviews (Backman \& Paresce~\cite{bac:par};
Vidal--Madjar \& Ferlet~\cite{v-m:f};
Artymowicz~\cite{art94}, \cite{art96}, \cite{art97}).  In the last review,
Artymowicz~(\cite{art97}) analyzed
three principal components of the $\beta$ Pictoris system: the star itself,
the circumstellar dust and the circumstellar gas.  Although there
is an improved determination of the distance towards $\bet$ from
the Hipparcos satellite (Crifo et al.~\cite{crifo}),
the main conclusions of the previous analyses remain in force.

In general, information on dust comes from measurements of extinction,
polarization, scattered light and thermal emission.
For scattering, the geometrical relation
between source, scatterer and observer is essential.  Whereas it is ill
determined in reflection nebulae and allows only a very rough
derivation of the grain properties, the geometrical configuration in the
circumstellar disc of $\beta$ Pic is clear and simple.  In the case of $\bet$
one observes:
\bi
\item [{\it i})] no circumstellar extinction (Crifo et al.~\cite{crifo});
\item [{\it ii})] very weak polarization of the star itself
(Tinbergen~\cite{tinb});
\item [{\it iii})] scattered light from the nearly edge--on  disc
(Smith \& Terrile~\cite{s:t}; Kalas \& Jewitt~\cite{k:j95};
Mouillet et al.~\cite{mou97})
and its polarization (Gledhill et
al.~\cite{gledh}; Wolstencroft et al.~\cite{wo95});
\item [{\it iv})] an infrared excess (Aumann et al.~\cite{aum})
which extends up to 1300$\mkm$ (Chini et al.~\cite{chi91});
there are also mid IR images (Lagage \&
Pantin~\cite{l:p}; Pantin et al.~\cite{p:l:art97}).
\ei

Many numerical calculations have been performed to explain the IR emission of
$\bet$ (see detailed discussion in Li \& Greenberg~\cite{li:gre98}).
Depending on the wavelength range considered, 
the observations were reproduced using compact or fluffy grains
with sizes ranging from smaller 1\,$\mkm$ to up to 1\,mm.

With respect to modelling the scattering and polarization of light
from the disc of $\beta$ Pic, Artymowicz et al.~(\cite{art89}),
Kalas \& Jewitt~(\cite{k:j95}, \cite{k:j96}) and
Pantin et al.~(\cite{p:l:art97}) considered scattering only at one wavelength.
They assumed either isotropic or anisotropic scattering without
computing the asymmetry factor $g$.  Scarrott et al.~(\cite{sca92}),
on the other hand, applied Mie theory and treated multi--wavelength scattering,
however, the polarization was only
calculated at the then available R--band.
Artymowicz~(\cite{art97}) reproduced the
observation in the V--band employing the empirical phase and polarization
function of zodiacal and cometary dust.

In this paper, we model scattering and polarization at all observed wavelengths
with particle cross sections computed from Mie theory.  As a result, we are able
to constrain the properties of the grains and to exclude certain choices of dust
models which were hitherto thought possible.

\section{Observations of polarization and colours}
\begin{figure}                                               %
\resizebox{8.8cm}{!}{\includegraphics{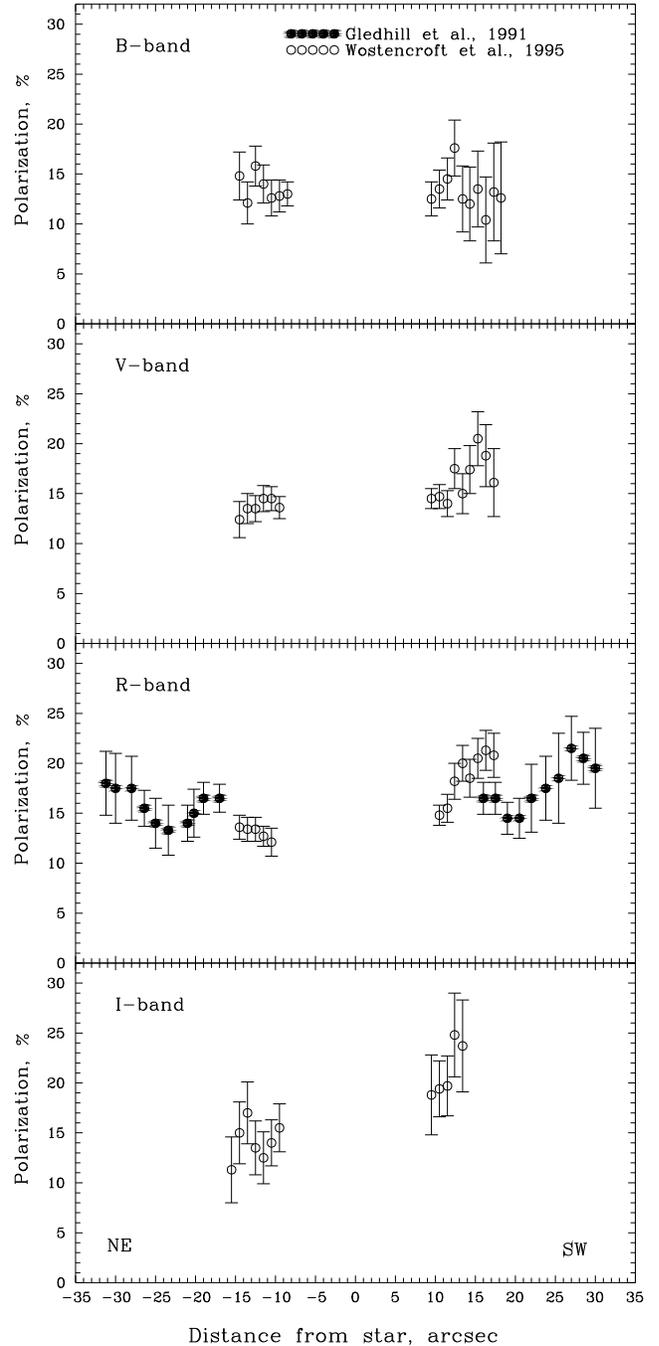}}
\caption[]{Results of the multi--wavelength measurements of the polarization
along the disc of $\bet$.}
\label{fig1}
\end{figure}

Imaging polarimetry of the $\bet$ disc was performed by
Gledhill et al.~(\cite{gledh})
in the R waveband and by Wolstencroft et al.(~\cite{wo95})
in the B, V, R and I
wavebands.  The polarization patterns are centro--symmetric and indicate that
$\bet$ itself is the illuminating source.  The observational data are collected
in Fig.~\ref{fig1} and suggest the asymmetry in the polarizing
behaviour between the SW and NE sides, especially in the I--band.

The wavelength dependence of polarization $P(\lambda)$
is typical of reflection nebulae and characterized by the increase of the
polarization degree with wavelength.  The increase is rather weak in
the NE side  and well pronounced in the SW side.

The polarization vectors are oriented perpendicular to the disc (Gledhill et
al.~\cite{gledh}; Wolstencroft~\cite{wo98}).  Tinbergen~(\cite{tinb}) included $\bet$ in the list of
zero polarization standard stars. Using a large diaphragm which included the
disc he found that the mean degree of polarization over the wavelength range
$0.4 - 0.7\,\mkm$ is $P = 0.020 \pm 0.008\%$.  This value is compatible with the
absence of material in front of the star under the assumption of a maximum
polarization efficiency ($P/A_V = 3$\%) and implies $A_V \simeq 0.006$\,mag.
The small polarization observed by Tinbergen~(\cite{tinb})
should be due to dust scattering in the disc.

\begin{figure}                                                %
\resizebox{8.7cm}{!}{\includegraphics{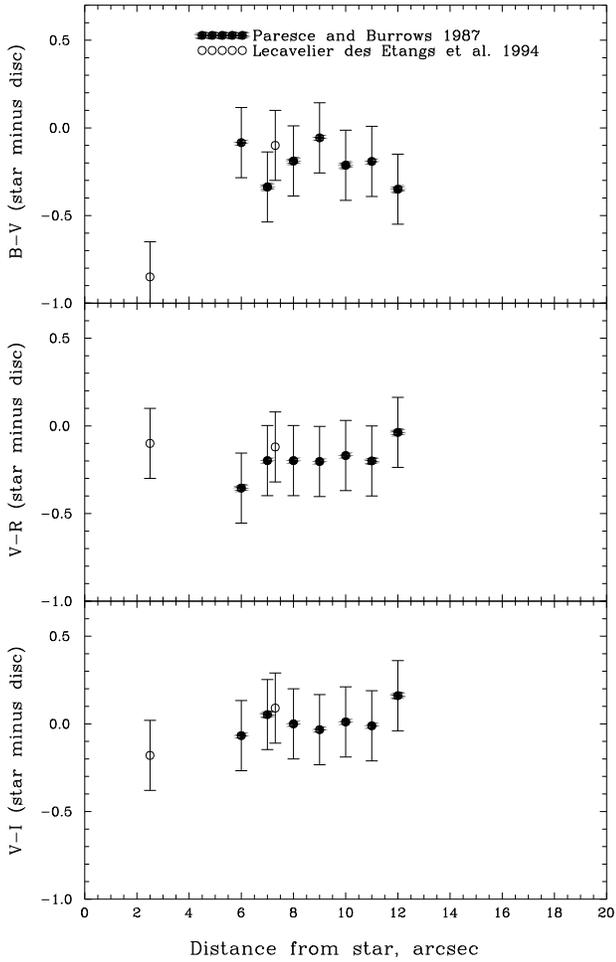}}
\caption[]{Observed colours along the disc of $\bet$.}
\label{fig2}
\end{figure}

In ordinary reflection nebulae, the colour of the scattered light is usually
bluer than the illuminating star at small offsets and redder at large distances
(Voshchinnikov~\cite{vo85}).  Unfortunately, the disc colours of $\bet$ were observed
only out to $\sim$12$''$ offset (Paresce \& Burrows~\cite{p:b}; Lecavelier des Etangs
et al.~\cite{lec}).  The data of Fig.~\ref{fig2} indicate that the disc has the same colour as
the star or is slightly redder.  The colours do not depend on position, the only
exception being the innermost point ($\varphi = 2\farcs5$).  It falls into the
central gap which is presently the subject of various speculations.

We thus conclude that the properties of the scattered light are not peculiar.
We therefore believe that the usual approach to the study of reflection
nebulae can also be applied to $\bet$.

\section{Modelling}
\subsection{Polarization mechanism and disc model}
The variation of the position angle and the degree of polarization in the disc
of $\bet$ speaks in favour of light scattering by spheres
or arbitrarily oriented non--spherical particles.
The disc is seen nearly edge--on. As it is optically thin,
we can choose a model of single scattering in an optically thin medium.
In this case, the radiation escaping
from the disc at the angular distance $\varphi$ from the star
is the integral along the line of sight over some range of scattering
angles ($\Theta_1, \, 180{\degr}-\Theta_1$), where
\be
\Theta_1 = \arcsin \left(\frac{\varphi}{\varphi_{\rm max}} \right)
\label{eq1}
\ee
and $\varphi_{\rm max}$ denotes the maximum angular size of the disc.
We will adopt $\varphi_{\rm max} = 45\farcs0$ (Kalas \& Jewitt~\cite{k:j95}).
With growing $\varphi$, the range of
scattering angles, which is always centered at $90{\degr}$,
becomes narrower.  A slightly tilted disc orientation ($\sim$10${\degr}$)
would not change the picture of light scattering significantly.

\subsection{Step 1: polarization diagrams vs scattering angle}
As a first step in the modelling, we compute the polarization and scattered
intensity of an elementary volume. In Mie scattering by spherical particles,
such polarization at a given wavelength $\lambda$ depends on the refractive
index of the particle, $m_{\lambda} = n_{\lambda} - k_{\lambda}i$, and its
radius $a$,
\be
P(m, a, \lambda, \Theta) = \frac{i_1(m, a, \lambda,
                           \Theta)-i_2(m,a,\lambda,\Theta)} {i_1(m, a, \lambda,
                           \Theta)+i_2(m,a,\lambda,\Theta)} \ .
\ee
Here $i_1$ and $i_2$ are the dimensionless intensities which determine the
radiation scattered perpendicular and parallel to the scattering plane,
respectively (van de Hulst~\cite{vdh}; Bohren \&  Huffman~\cite{b:h}).

\begin{figure}                                            %
\resizebox{8.2cm}{!}{\includegraphics{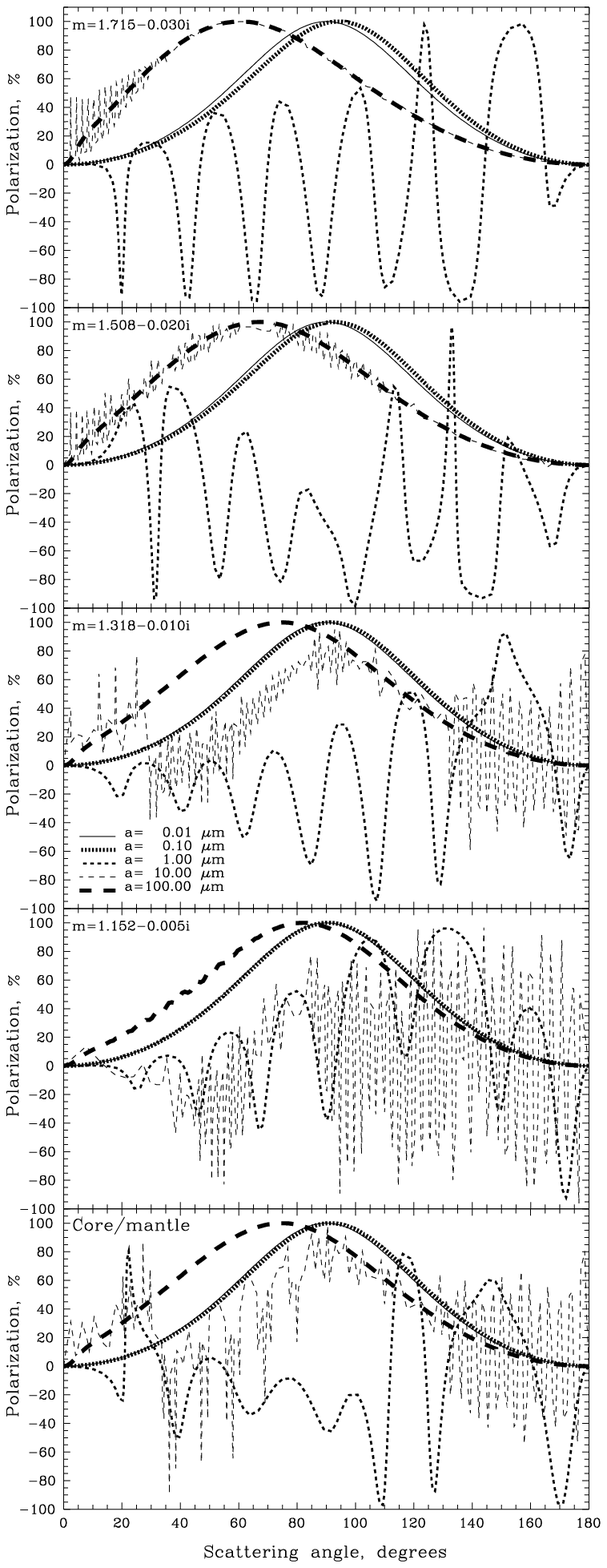}}
\caption[]{
The degree of linear polarization for various kinds of particles.  Their
composition and  refractive indices are given in Table~\ref{tab1}.
The peak of the curve for $a=0.1\,\mkm$ is always slightly shifted to
larger values of scattering angle relative to those for $a=0.01\,\mkm$.}
\label{fig3}
\end{figure}

Figure~\ref{fig3} shows the polarization diagrams at $\lambda=0.70\,\mkm$
(R--band) for which the observational database is largest (Fig.~\ref{fig1}).
The particles considered are a mixture of silicate and ice.
They could be porous,
i.e.~contain some volume fraction of vacuum. Porous grains have been
discussed many times for the disc of $\bet$
(see Artymowicz~\cite{art97}; Li \& Greenberg~\cite{li:gre98}).
The refractive indices used are specified
in Table~\ref{tab1}. Note that we chose the volume fractions of the grain
constituents in such a way that several of the refractive indices
in Table~\ref{tab1} are identical, although they refer to grains of
different chemical composition.  The
cross sections of grains with a silicate core
and an ice mantle were computed from G\"uttler's~(\cite{gu52}) theory
for two--layered spheres.
\begin{table*}
\caption[]{Grain refractive indices $m_R$ at $\lambda=0.70\,\mkm$ }
\begin{flushleft}
\begin{tabular}{ll}
\noalign{\smallskip}
\hline
\noalign{\smallskip}
Refractive index & ~~Particle composition  \\
\noalign{\smallskip}
\hline
\noalign{\smallskip}
$1.715 - 0.030i$ & ~~compact silicate grains   \\
$1.508 - 0.020i$ &
   $\left\{  \begin{array}{l}
   {\rm composite \ \mbox{grains:} \ mixture \ of \ 50\% \ silicate} + 50\% \ {\rm ice}  \\
   {\rm porous \ \mbox{grains:} \ mixture \ of \ 72\% \ silicate} + 28\% \ {\rm vacuum}
   \end{array} \right.$ \\
\noalign{\smallskip}
$1.310 - 0.010i$ &
   $\left\{  \begin{array}{l}
   {\rm compact \ grains \ of \ dirty \ ice}  \\
   {\rm porous \ \mbox{grains:} \ mixture \ of \ 45\% \ silicate} + 55\% \ {\rm vacuum}
   \end{array} \right.$ \\
\noalign{\smallskip}
$1.152 - 0.005i$ &
   $\left\{  \begin{array}{l}
   {\rm porous \ \mbox{grains:} \ mixture \ of \ 50\% \ ice} + 50\% \ {\rm vacuum} \\
   {\rm porous \ \mbox{grains:} \ mixture \ of \ 24\% \ silicate} + 76\% \ {\rm vacuum}
   \end{array} \right.$ \\
$1.715 - 0.030i$/$1.310 - 0.010i$ & ~~core--mantle grains: silicate core $+$ ice
mantle   \\
\noalign{\smallskip}
\hline
\noalign{\smallskip}
\end{tabular}

Optical constants for silicate are from Draine~(\cite{draine85}), for dirty ice from
Greenberg~(\cite{gre68}). \\
The optical constants for composite and porous grains were
obtained from the Bruggeman rule (Bohren \& Huffman~\cite{b:h}).
\end{flushleft}
\label{tab1}
\end{table*}

It is important to note that all polarization diagrams of Fig.~\ref{fig3}
are similar and independent of the particle composition and structure.
Small grains belong to the Rayleigh domain where the polarization
has the well known bell--like shape with the maximum at $\Theta = 90{\degr}$.
Polarization diagrams for very large
spheres are also simple.  They resemble smooth curves with the maximum at
$\Theta \approx 70{\degr}$.  In both cases, the polarization does not change
sign and reaches $\sim$100\%.  For particles of intermediate size, polarization
reverses sign repeatedly and a ripple--like structure is seen.

This behaviour of the curves $P(\lambda)$ in Fig.~\ref{fig3} reflects the general
principles of light scattering by small particles and does not principally
change for non--spherical bodies, like spheroids, cylinders, bispheres, or
fluffy aggregates arbitrarily aligned in space (Mishchenko et al.~\cite{mi96};
Kolokolova et al.~\cite{kol}); Lumme et al.~\cite{lumme}).  Moreover, Lumme et al.~(\cite{lumme})
demonstrated that Mie theory of spheres reproduces the polarization diagrams for
complex particles rather well, except sometimes for forward and backward
scattering.

The polarization measurements in $\bet$ were made at offsets from
$\varphi = 8\farcs5$ to $31\farcs2$ (see Fig.~\ref{fig1}).
According to Eq.~(\ref{eq1}), the scattering
angles vary in the limits from ($11{\degr}-169{\degr}$)
to ($44{\degr}-136{\degr}$).  With these limits,
we conclude with the aid of Fig.~\ref{fig3}:
{\it the polarimetric observations of $\beta$ Pic cannot be explained by
light scattering of very small {\rm (}$< 0.1\mkm${\rm )} or very large
{\rm (}$> 10\mkm${\rm )} particles only
because they produce the polarization too high for the scattering angle
range.}

\subsection{Step 2: polarization diagrams vs particle size}
As the second step of the modelling procedure we include averaging
along the line of sight (scattering angles) at fixed angular
distance $\varphi$. We adopt that the number density of dust
grains in the disc has a power--law distribution
\bea
  n_{\rm d}(r) =
  \left\{ \begin{array}{ll}
            0,  & \mbox{$r < r_0$}, \\
         n_0 \left(\frac{r}{r_0} \right)^{-s},
           & \mbox {$ r_0 \leq r < r_{\rm out}.$}
        \end{array}
         \right.
           \label{dens}
\eea
Here, $r_0$ is the radius of the central hole where the density is equal to
$n_0$ and $r_{\rm out}$ is the outer radius of the disc.  The latter
is given by $r_{\rm out}=D \varphi_{\rm max} = 19.28 \cdot 45 \approx 870$\,AU,
assuming a distance $D = 19.28$\,pc (Crifo et al.~\cite{crifo}).
The central cavity reaches out to $\varphi \la\varphi_0 \approx 6\farcs0$
(Artymowicz et al.~\cite{art89}).\footnote{The adaptive optics observations
made by Mouillet et al.~(\cite{mou97}) show that the scattered light is present
at angular distances up to $\varphi \approx 1\farcs5$.
Perhaps, this is the light from the outer disc scattered in almost forward
and backward directions. However, because the polarization observations
are made outside of this radius, this does not impact on the result, given
in the single scattering case.}
With these numbers, $r_0 \approx 120$\,AU and the ratio of the inner
to outer radius equals
$\varphi_0/\varphi_{\rm max} = r_0/r_{\rm out} \approx 0.13$.

The polarization degree can be expressed as
\be
P(m,a,\lambda,\varphi,s) \ = \ \frac{<I_1> - <I_2>} {<I_1> + <I_2>} \ ,
\ee

\be
<I_j> \ = \ \frac{{\cal K}}{\varphi^{s+1}} \int_{\Theta_1}^{\pi-\Theta_1}
              i_j(m,a,\lambda,\Theta) \sin^s\Theta \, {\rm d}\Theta \ ,
\ee
where the intensities $I_j$ $(j=1,2)$ depend on $m,a,\lambda,\varphi,s$, and
${\cal K}$ is a constant.  If $\varphi < 6\farcs0$ and $\sin \Theta > \frac{1}
{0.13} \cdot \frac{\varphi}{\varphi_{\rm max}}$, the line of sight intersects
the hole.

\begin{figure}                                                %
\resizebox{8.8cm}{!}{\includegraphics{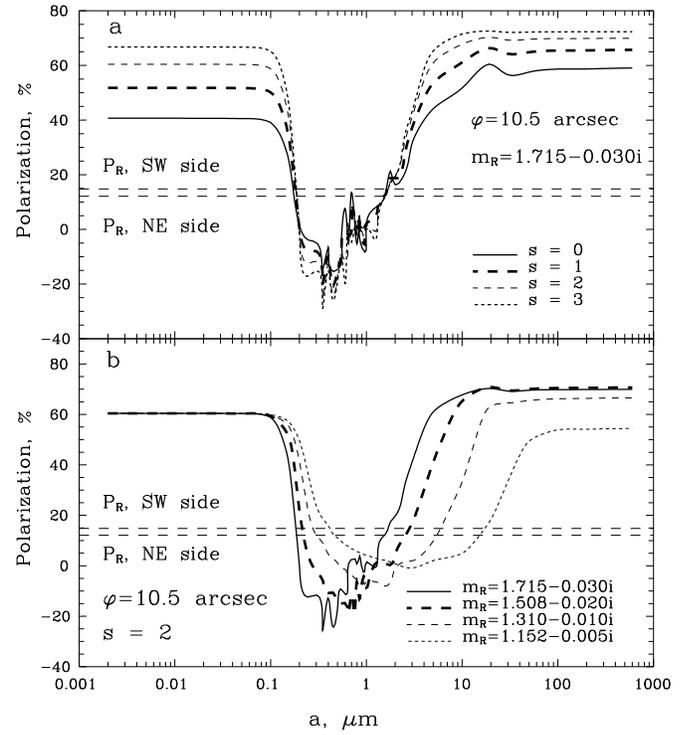}}
\caption[]{
({\bf a}) Upper panel: polarization vs particle size at
the angular distance $\varphi=10
\farcs5$ for various exponents, $s$, of the dust density distribution.
The grains are compact silicates with $m_R$ as indicated.
The horizontal dashed lines show the observed polarization $P_R$
in the NE and SW sides.
({\bf b}) Lower panel: as ({\bf a}), but now for a fixed
density distribution ($s=2$)
and particles of various refractive indices.  }
\label{fig4}
\end{figure}

Examples of polarization diagrams are shown in Fig.~\ref{fig4} at the offset $\varphi
=10\farcs5$ for different types of particles and various values of $s$.  The
polarization degree observed in the R--band at this offset angle in the NE and SW
extensions is shown by horizontal dashed lines.  From this figure we again
conclude that grains of intermediate size have to be included into
consideration to explain the observed polarization.
Interestingly, the polarization observed at
given offset may be explained for any density distribution of particles: as
evident from Fig.~\ref{fig4}a (the parameter $s$ determines
only the polarization level at low and large size limits).

\begin{figure}                                                %
\resizebox{8.8cm}{!}{\includegraphics{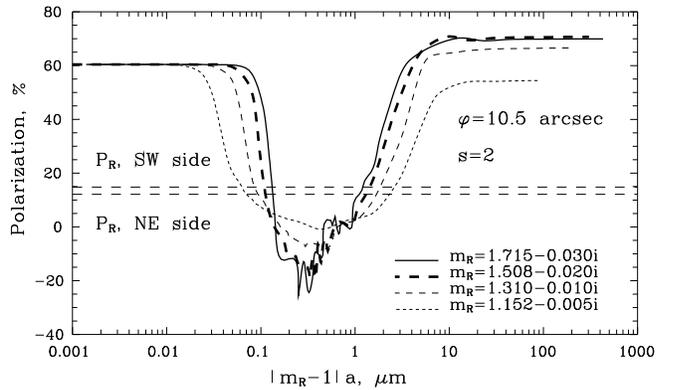}}
\caption[]{
As Fig.~4b, but with polarization plotted vs the parameter $|m_R-1|a$.}
\label{fig5}
\end{figure}

The curves plotted in Fig.~\ref{fig4}b look even more similar
if we use the product $|m-1|a$ as an argument.
That this is so is well
known for the extinction efficiency (e.g.~Greenberg~\cite{gre68}).
The corresponding polarization diagrams are shown in Fig.~\ref{fig5}.
Thus, considering light scattering in the continuum,
the effects of refractive index and particle size
cannot be separated.  In other words, {\it we can only estimate $|m-1|a$, the
product of refractive index times particle size, from observations at one
wavelength}.

In more realistic models, we must also average over the size distribution
$n_{\rm d}(a)$.  Afterwards the spikes in polarization seen
in Fig.~\ref{fig3} disappear. We use the power--law
\be
     n_{\rm d}(a) \sim  a^{-q}  \label{size_dis}
\ee
\begin{figure}                                                %
\resizebox{8.8cm}{!}{\includegraphics{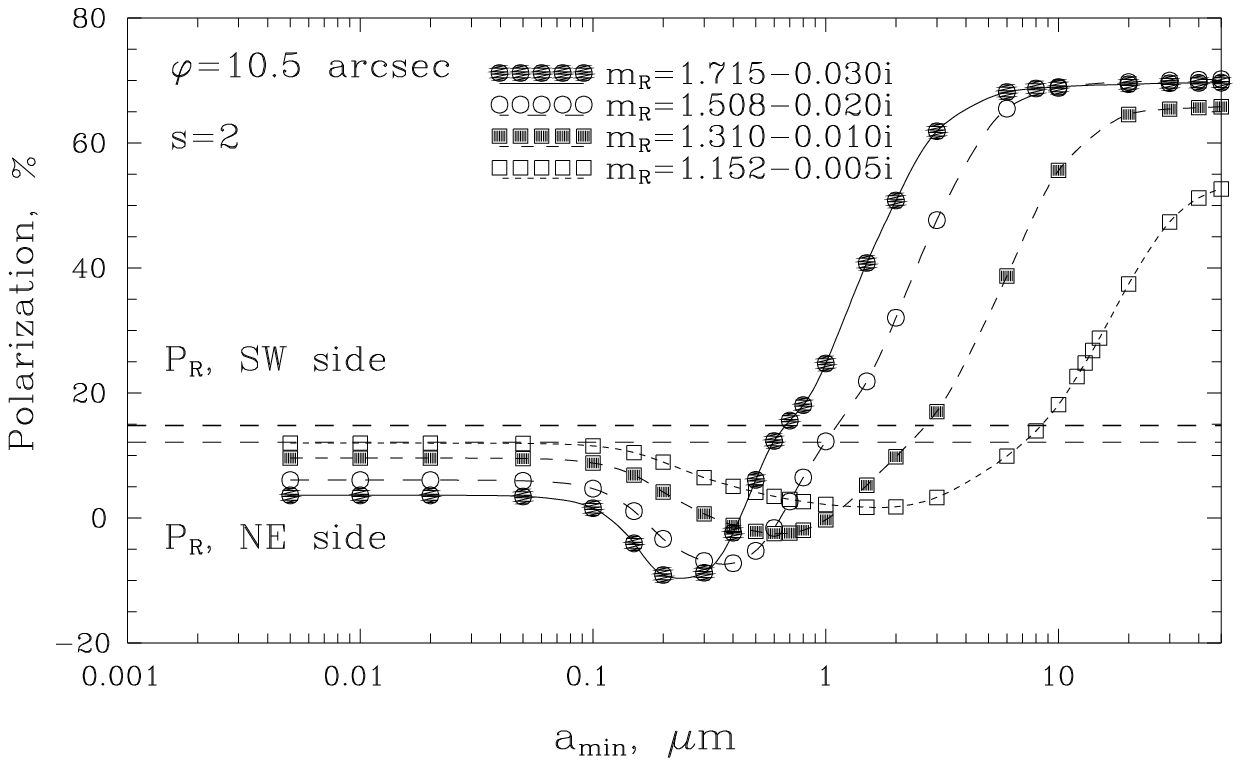}}
\caption[]{
Polarization vs $a_{\rm min}$ at the angular distance $\varphi=10\farcs5$.  The
upper size limit equals $a_{\rm max} = 100\,\mu$m and the exponent of the size
distribution $q=3.5$.}
\label{fig6}
\end{figure}                                                %
%
\begin{figure}                                                %
\resizebox{8.8cm}{!}{\includegraphics{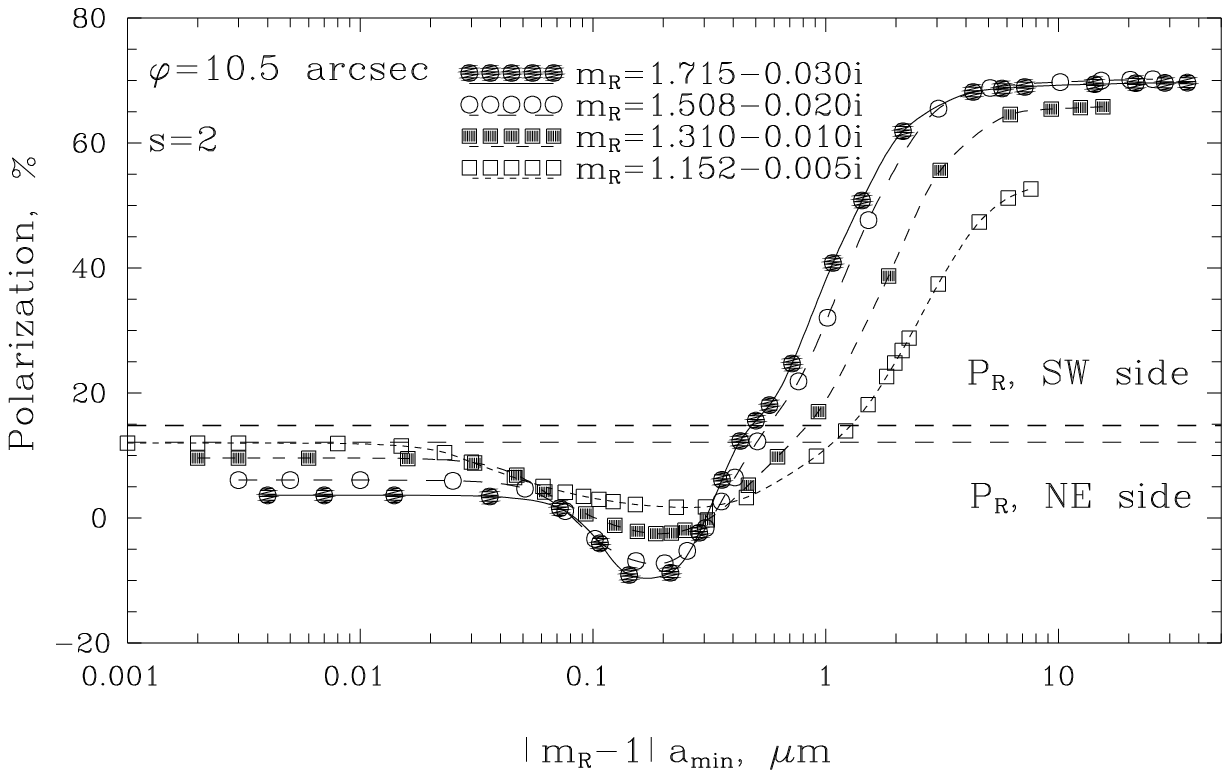}}
\caption[]{
As Fig.~6, but with polarization vs the parameter $|m_R-1|a_{\rm min}$.}
\label{fig7}
\end{figure}                                                %
with minimum and maximum radii $a_{\rm min}$ and $a_{\rm max}$, respectively.
Figures~\ref{fig6} and \ref{fig7} are analogous to Fig.~\ref{fig4}
and \ref{fig5}, but include an average over the
size distribution. The upper radius and the exponent are kept fixed
($a_{\rm max}=100\,\mu$m$, q=3.5$) and $a_{\rm min}$ is being varied.

The changes of $a_{\rm min}$ are most important with respect to polarization.
The increase or decrease of the maximum size has no influence if $|m-1|a_{\rm
max} \ga 10$.  So if very large particles are present in the disc of $\beta$
Pic, they are not visible as scatterers at waveband R.  Changes in $q$ do not
affect the overall picture and will be discussed in Section 3.6.
Figures~\ref{fig6} and \ref{fig7} contain plots for all refractive indices listed
in Table~\ref{tab1}.
The figures show that {\it in order to explain the observed polarization one
needs to keep $a_{\rm min}$ in the range $0.6\,\mkm \la$ $a_{\rm min}$
$\la 10\,\mkm$}.  However, very porous grains with small values of $a_{\rm min}$
may, in principle, be used to model the observations as well.

\subsection{Step 3: polarization vs angular distance from star}
\begin{figure}                                                %
\resizebox{8.8cm}{!}{\includegraphics{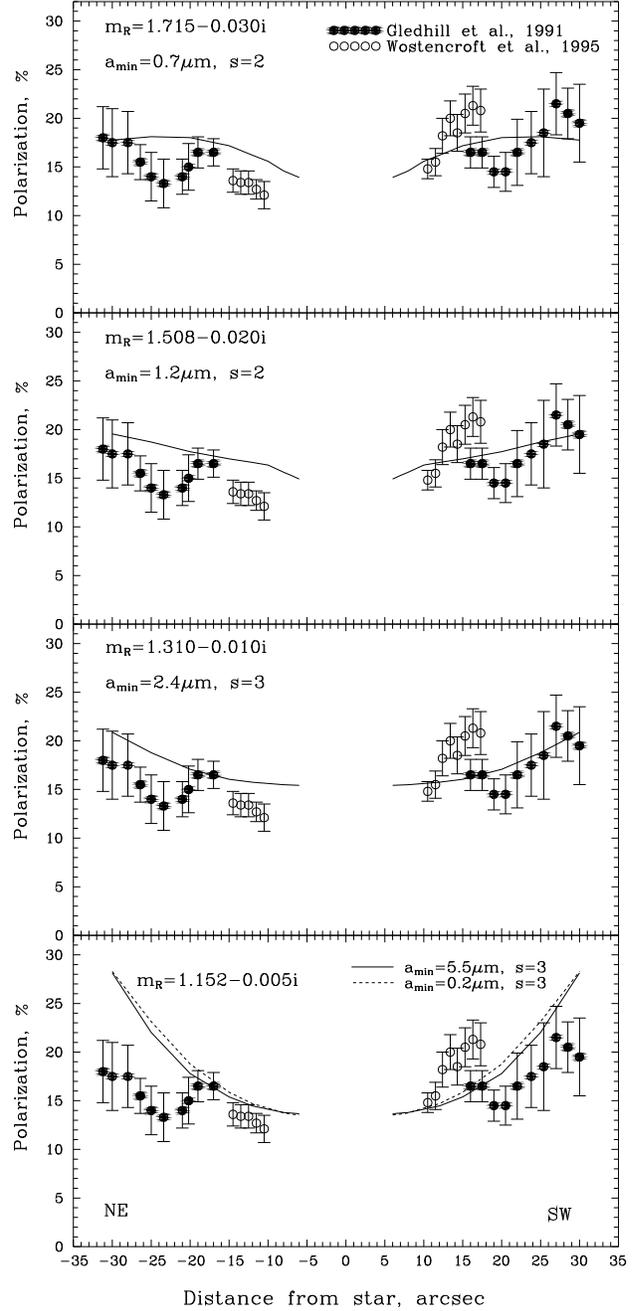}}
\caption[]{
Polarization in the disc of $\bet$ as a function of the angular offset for
the R--band.  Solid lines present models with different refractive indices
(see Table~\ref{tab1}).  In all calculations, $a_{\rm max}=100\,\mkm$
and $q=3.5$.  The other model parameters ($a_{\rm min}$  and $s$) are
shown in the panels.}
\label{fig8}
\end{figure}

In the next step, we model the angular distribution of polarization using the
minimum cut--off in the size distribution as estimated from Fig.~\ref{fig6}.  The curves
in Fig.~\ref{fig8} demonstrate that the same observations may be satisfactorily fit with
particles of different composition by changing only the parameters $a_{\rm min}$
and $s$.\footnote{Actually, the choice of the parameter $s$ is determined by the
radial brightness distribution of the scattered light, $I(\varphi)$.
For the outer parts of the disc of $\bet$ ($\varphi > 6\farcs0$), $I(\varphi)
\propto \varphi^{- \gamma}$ with $\gamma=3.6-4.3$ (Kalas \& Jewitt~\cite{k:j95}).
The values of $\gamma$ for the models considered are given in Table~\ref{tab2}.}
Note that the variations of $s$ and $a_{\rm min}$ influence the  polarization
mainly at small and large offset angles.  (We adopted $a_{\rm max}=100\,\mkm$
and $q=3.5$.)  Therefore, {\it the polarimetric observations of the extended
sources (nebulae, circumstellar shells, galaxies) at one wavelength do
not allow to determine the chemical composition of particles}.  Numerous other
sets of refractive indices and size parameters may also explain the
observations.  As an example, let us consider the model of Scarrott et
al.~(\cite{sca92}).  They reproduced the angular dependence of the scattered light and
polarization in the R--band observed by Gledhill et al.~(\cite{gledh}) with silicate
particles of $m_R=1.65-0.05i$ and $a_{\rm min}=0.01\,\mkm$, $a_{\rm max}=3\,\mkm$,
$q=4.0$, $s=2.75$.  However, the disc colours in their model are too red
compared to those observed (from
$-0\fm72$ in (B--V)$_{\star - {\rm disc}}$ to $-1\fm24$ in
(V--I)$_{\star - {\rm disc}}$).

\subsection{Step 4: polarization vs wavelength}
With the parameters found from modelling of the polarization in the R--band
(see Fig.~\ref{fig8}), we calculated the wavelength dependence of
polarization at the offset distance $\varphi=10\farcs5$.  We also checked
the colour excesses given by the
corresponding model (see Table~\ref{tab2}) and compared
them with observations.
\begin{figure}                                                %
\resizebox{8.8cm}{!}{\includegraphics{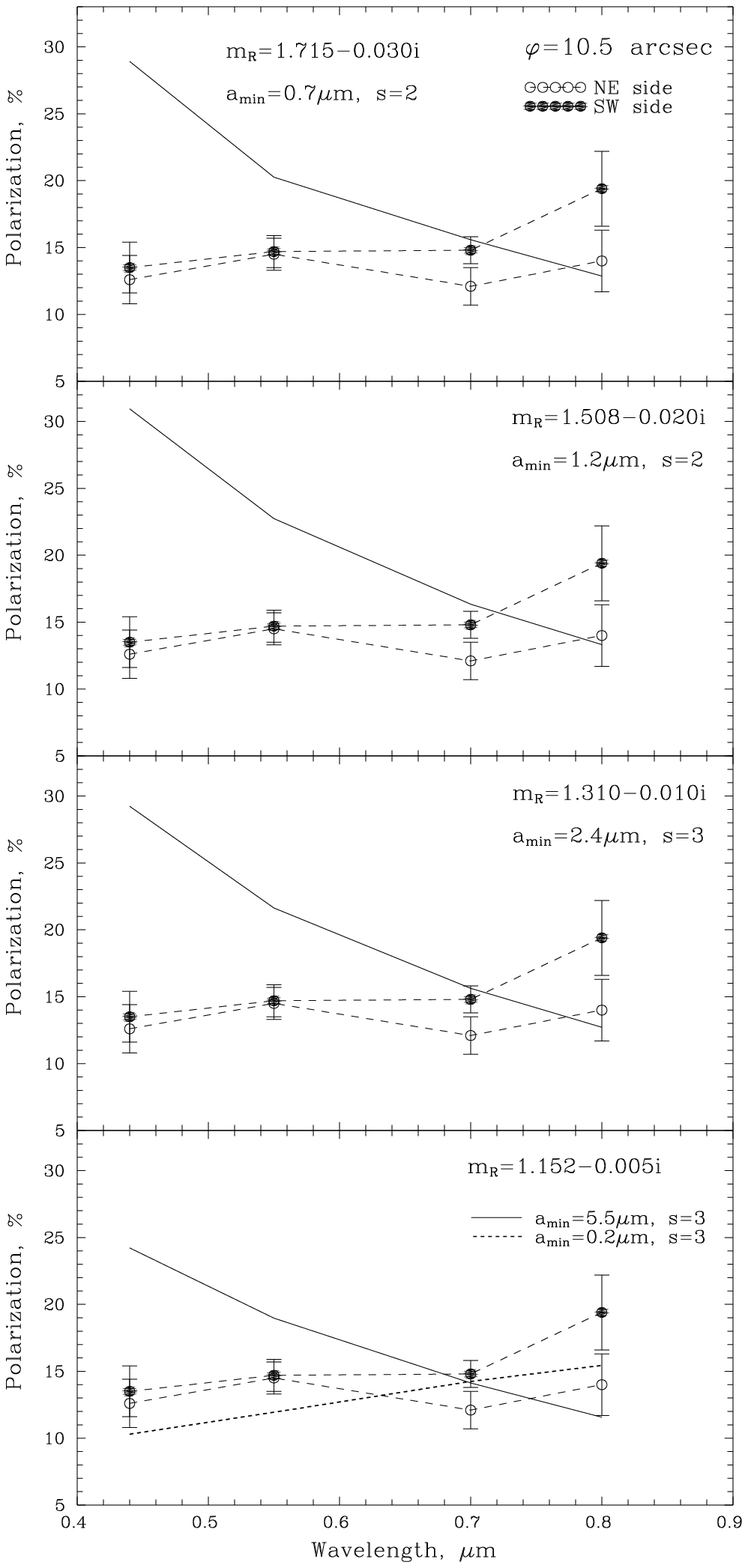}}
\caption[]{
The wavelength dependence of polarization in the disc of $\bet$ at the angular
distance $\varphi=10\farcs5$.  The observations in the SW and NE sides and their
errors are plotted by dashed lines.  The parameters of the model are the same as
in Fig.~8.}
\label{fig9}
\end{figure}                                                %

The results are shown in Fig.~\ref{fig9} and Table~\ref{tab2}.  It is clear
that despite our thorough procedure and the successful fits of
the changes in polarization and
brightness in the R--band with angle $\varphi$ (see Fig.~\ref{fig8}
and Table~\ref{tab2}), the
dependence $P(\lambda)$ cannot be well explained for any refractive index.
From Table~\ref{tab2} it also follows that the colour index V--I is crucial
for the choice of the model.  The colours in the models~1~--~4 are too red
and these models must be rejected.  In principle, the correct
polarization behaviour (i.e.~its growth
with wavelength) can be reproduced for all refractive indices of
Table~\ref{tab1} if the minimum size $a_{\rm min}$ is lowered.  But even
if we were to fit the dependence $P(\lambda)$, very small particles would
yield unacceptably blue colours.
\begin{table*}
\caption[]{
Observational and theoretical colour excesses in the disc of $\bet$ at
$\varphi=10\farcs5$.   }
\begin{flushleft}
\begin{tabular}{lllllll}
\noalign{\smallskip}
\hline
\noalign{\smallskip}
Colour excess & ~Observations & Model 1 & Model 2 & Model 3 & Model 4 & Model 5 \\
\noalign{\smallskip}
\hline
\noalign{\smallskip}
(B--V)$_{\star - {\rm disc}}$ & $-0\fm21 \pm 0\fm20$ &$-0\fm217$&$-0\fm213$& $-0\fm221$& $-0\fm228$& $\phantom{-}0\fm034$\\
(V--R)$_{\star - {\rm disc}}$ & $-0\fm17 \pm 0\fm20$ &$-0\fm200$&$-0\fm226$& $-0\fm228$& $-0\fm252$& $\phantom{-}0\fm061$\\
(V--I)$_{\star - {\rm disc}}$ & $\phantom{-}0\fm01 \pm 0\fm20$ &$-0\fm313$&$-0\fm344$& $-0\fm354$& $-0\fm400$& $\phantom{-}0\fm105$\\
\noalign{\smallskip}
\hline
\noalign{\smallskip}
\end{tabular}

Model 1: $m_R=1.715-0.030i$, $a_{\rm min}=0.7\,\mkm$, $s=2$
         ($\gamma=3.4$, $\Lambda=0.67$, $g=0.80$). \\
Model 2: $m_R=1.508-0.020i$, $a_{\rm min}=1.2\,\mkm$, $s=2$
         ($\gamma=3.4$, $\Lambda=0.66$, $g=0.86$). \\
Model 3: $m_R=1.310-0.010i$, $a_{\rm min}=2.4\,\mkm$, $s=3$
         ($\gamma=4.2$, $\Lambda=0.67$, $g=0.91$). \\
Model 4: $m_R=1.152-0.005i$, $a_{\rm min}=5.5\,\mkm$, $s=3$
         ($\gamma=4.4$, $\Lambda=0.67$, $g=0.96$). \\
Model 5: $m_R=1.152-0.005i$, $a_{\rm min}=0.2\,\mkm$, $s=3$
         ($\gamma=4.4$, $\Lambda=0.89$, $g=0.82$). \\
All models have $a_{\rm max} = 100\,\mu$m and $q=3.5$.\\
$\gamma$ --  exponent in the brightness distribution,
$\Lambda$ -- single scattering albedo, $g$ -- asymmetry factor.
\end{flushleft}
\label{tab2}
\end{table*}

{\it So in order to explain all colour and polarization observations we need to
reduce the lower limit of the size distribution, $a_{\rm min}$, and the
refractive index} $m$.

\subsection{Step 5: polarization and colours: whole picture}
\begin{figure}                                           %
\resizebox{8.8cm}{!}{\includegraphics{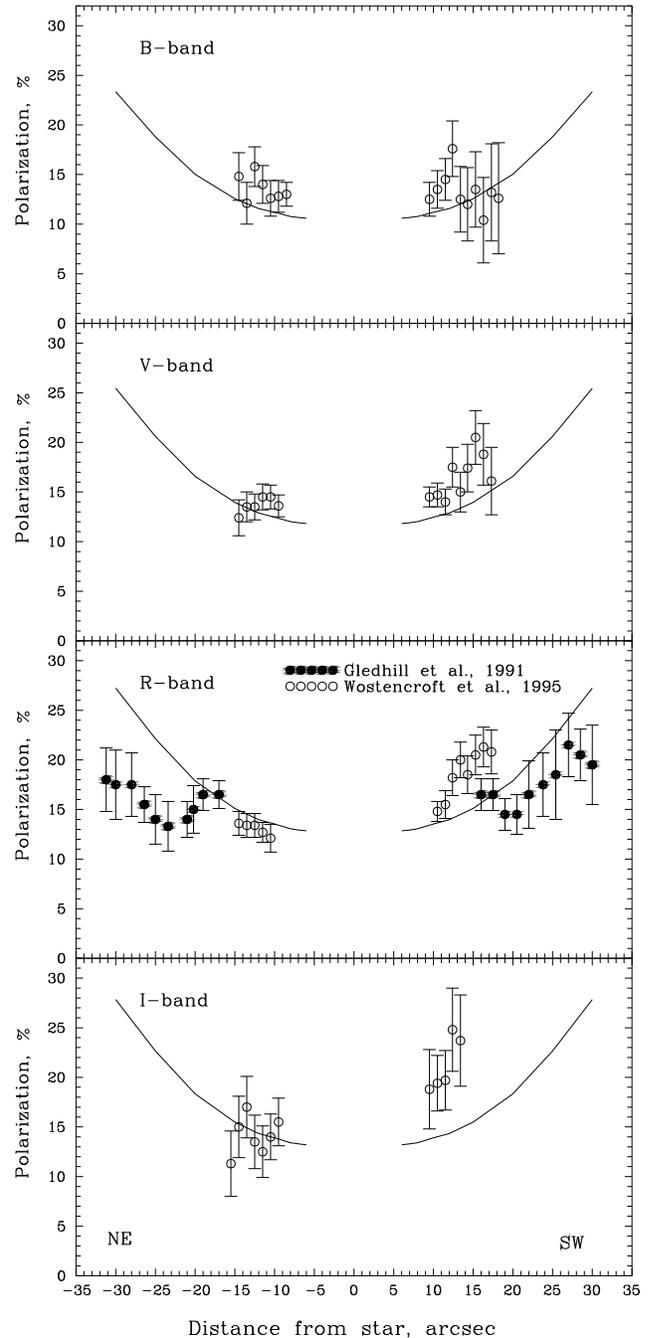}}
\caption[]{
The polarization observed in the disc of $\bet$.
The solid lines correspond to the
theoretical models for which $m_R = 1.152 - 0.005i$, $a_{\rm min}=0.15\,\mkm$,
$a_{\rm max}=100\,\mkm$, $q=3.2$ and $s=3$.}
\label{fig10}
\end{figure}
\begin{figure}                                                %
\resizebox{8.8cm}{!}{\includegraphics{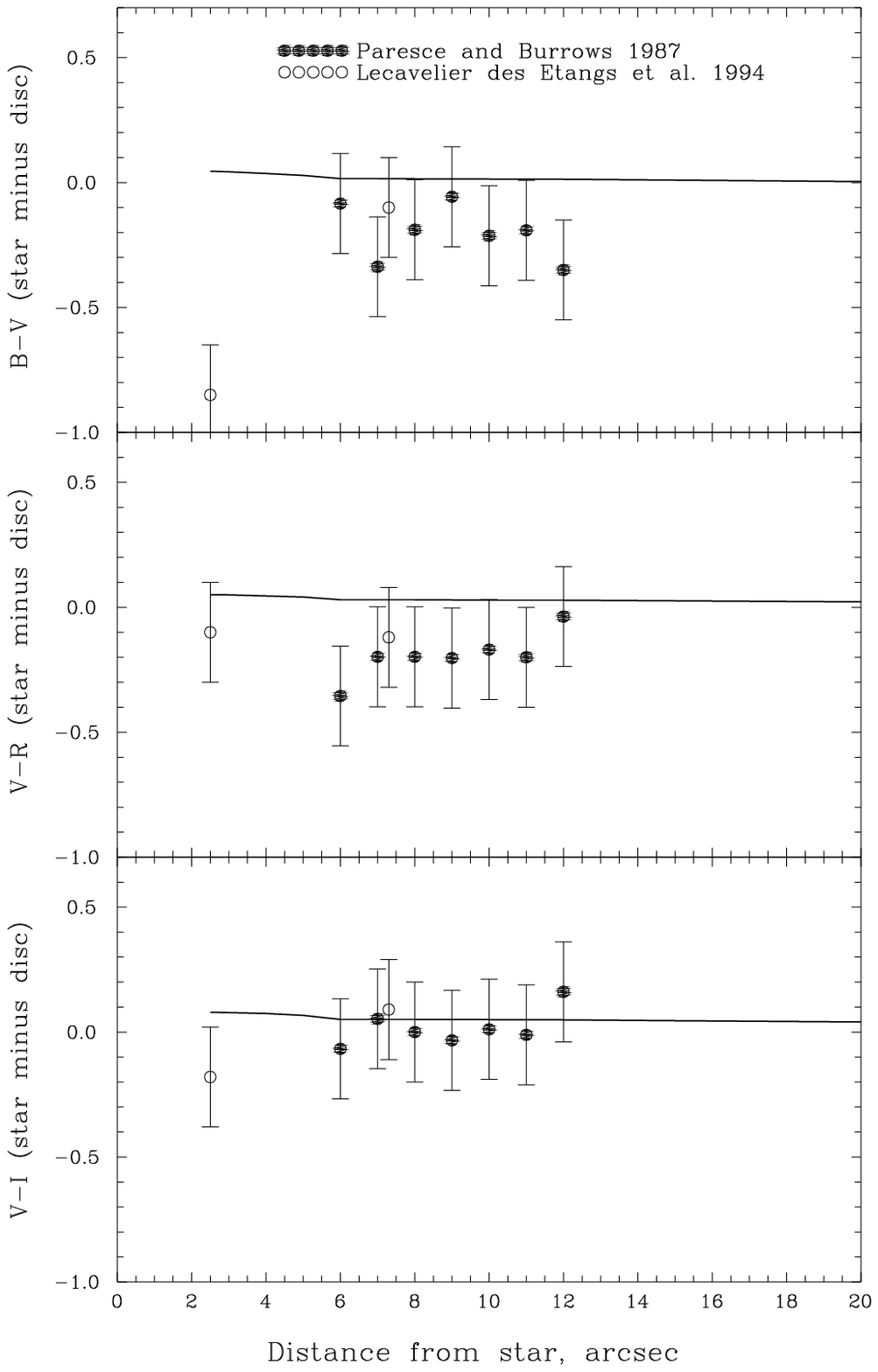}}
\caption[]{
The colour excesses observed in the disc of $\bet$.
The theoretical model has the
same parameters as in Fig.~10.}
\label{fig11}
\end{figure}

                                                %
Finally, we suggest a fit to the angular dependence of the polarization
(in all wavebands) and the colour excesses (see Figs.~\ref{fig10} and \ref{fig11}).
In order to improve the fits, we varied the parameters $a_{\rm min}$ and $q$
around the values given in Table~\ref{tab2} for model 5.
Unfortunately, there remains some problem for the
position with the minimum distance $\varphi = 2\farcs5$.  The colour
index B--V can be modeled only if the grains have very narrow size
distribution around $a \sim 15\,\mkm$, but with such grains the
polarization at observed offsets becomes too large (see Fig.~\ref{fig4}).

\section{Concluding remarks}
All models presented here were constructed with a small number of parameters
which are rather well determined.  Our favourite model presented
in Figs.~\ref{fig10}, \ref{fig11} has parameters:

$m_R = 1.152 - 0.005i$,

$a_{\rm min}=0.15\,\mkm$,

$a_{\rm max}=100\,\mkm$,

$q=3.2$,

$s=3$. \\ 
The model fits reasonably well the observed angular dependence of
polarization (in four wavebands), three colour excesses and
the radial brightness distribution.  The
calculated values of the exponent in the brightness distribution, single
scattering albedo and asymmetry factor in the R--band are:

$\gamma$ = 4.4,

$\Lambda$ = 0.88,

$g$ = 0.79. \\
Still better fits could be obtained assuming different dust properties
on either side of the disc.  Previous estimates of the grain albedo
in the disc of $\bet$ based on visible/IR models gave
$\Lambda = 0.5 \pm 0.2$ (Artymowicz~\cite{art97}), which is smaller
than our value for very porous grains.
But Artymowicz et al.~(\cite{art89}) also derived $\Lambda\simeq 0.9$.

As usual, the albedo $\Lambda$ of the grains was estimated
independently of the scattering phase function.  Evidently, the latter
cannot be isotropic in the visual because so far all suggested grain
sizes imply forward scattering.  The value of
our model ($g \approx 0.8$) is larger than previous ones.  Using the
Heyney--Greenstein phase function, Kalas \& Jewitt (\cite{k:j95})
found that the brightness distribution observed in the disc of $\bet$
in the R--band may be reproduced with $0.3 \leq |g| \leq 0.5$.
Note that mirror particles (with
$g<0$) used by Kalas \& Jewitt~(\cite{k:j95}, \cite{k:j96}) are astronomical
nonsense because no grain material (with the exception of very small
iron particles) gives at visual
wavelengths a negative asymmetry parameter.  Even pure conductors
($m= \infty $) have $g \approx -0.2$ only (van de Hulst~\cite{vdh}).

In our favourite model (see Figs.~\ref{fig10}, \ref{fig11}) we only specify
the refractive index (see Table~\ref{tab1}).
It was taken arbitrarily and is not motivated from the physical point of view.
From other side, the porous particles may be easily kept in the shell because
the radiation pressure acting on them is smaller:
we found that for silicate grains of radii $a \ga 2.0 \,\mkm$
with 76\% porosity (Table~\ref{tab1}) the radiation pressure
force is smaller than gravitational force.

Very porous aggregates were adopted by Li \& Greenberg~(\cite{li:gre98})
to explain the IR observations.  Mie theory can, of course, not be
applied in such cases. Although Li \& Greenberg used a standard mixing rule
to treat the fluffy aggregates, the underlying theory has not been developed
for particles with sizes larger than the wavelength nor for computing
the scattering of light.

We modeled scattering and polarization in the disc of $\bet$ at all wavelengths
where it has been observed.  Because the properties of the scattered light are
not peculiar, we could compute particle cross sections from Mie theory.  Our
models exclude the possibility that the grains are compact spheres, or that they
are all very small or all very large grains.  Instead, the particles
must be rather porous (degree of porosity $\ga 50\,\%$) and have
a lower cut--off in the size
distribution as small as $a_{\rm min} \sim 0.2 \, \mkm$.
This limit is larger than the mean size of interstellar grains.
Our model can be further verified by computing the resulting IR fluxes
and check how well they agree with observations.

\begin{acknowledgements}
The authors are thankful to Vladimir Il'in,  Nicolas Mauron and
an anonymous referee for helpful comments.  NVV wishes to
thank for hospitality Max--Planck--Institut f\"ur Radioastronomie where this work
was begun.  The work was partly supported by grants of
the program ``Astronomy'' of the government of the Russian Federation,
the program ``Universities of Russia -- Fundamental Researches''
(grant N~2154) and the Volkswagen Foundation.
\end{acknowledgements}


\end{document}